\def\build#1_#2^#3{\mathrel{
\mathop{\kern0pt#1}\limits_{#2}^{#3}}}
\def\ga{\mathrel{\mathchoice {\vcenter{\offinterlineskip\halign{\hfil
$\displaystyle##$\hfil\cr>\cr\sim\cr}}}
{\vcenter{\offinterlineskip\halign{\hfil$\textstyle##$\hfil\cr>\cr\sim\cr}}}
{\vcenter{\offinterlineskip\halign{\hfil$\scriptstyle##$\hfil\cr>\cr\sim\cr}}}
{\vcenter{\offinterlineskip\halign{\hfil$\scriptscriptstyle##$\hfil
\cr>\cr\sim\cr}}}}}
\def\la{\mathrel{\mathchoice {\vcenter{\offinterlineskip\halign{\hfil
$\displaystyle##$\hfil\cr<\cr\sim\cr}}}
{\vcenter{\offinterlineskip\halign{\hfil$\textstyle##$\hfil\cr<\cr\sim\cr}}}
{\vcenter{\offinterlineskip\halign{\hfil$\scriptstyle##$\hfil\cr<\cr\sim\cr}}}
{\vcenter{\offinterlineskip\halign{\hfil$\scriptscriptstyle##$\hfil
\cr<\cr\sim\cr}}}}}

\def\bar{\overline}
\def\ben{\begin{equation}}
\def\be#1{\begin{equation}\label{eq:#1}}
\def\ee{\end{equation}}
\def\EC#1{(\ref{eq:#1})}
\def\reff{\par\noindent\hangafter=1\hangindent=1cm}

\documentstyle[11pt,fleqn]{article}
\def\la{\mathrel{\mathpalette\fun <}}
\def\ga{\mathrel{\mathpalette\fun >}}
\def\fun#1#2{\lower3.6pt\vbox{\baselineskip0pt\lineskip.9pt
        \ialign{$\mathsurround=0pt#1\hfill##\hfil$\crcr#2\crcr\sim\crcr}}}

\def\ksm{{\rm \,km\,sec^{-1}\,Mpc^{-1}}}
\def\gyr{{\rm \,Gyr}}
\def\hmpc{{\rm \,h^{-1} Mpc}}
\def\mpc{{\rm \,Mpc}}
\def\om{\Omega_{\rm m}}
\def\ov{\Omega_{\rm void}}
\def\hl{H_{\rm L}}

\def\dbi{\delta_0}
\def\dbc{\bar{\delta}(\chi)}
\begin{document}
\begin{titlepage}
\null\vspace{-62pt}
\vspace{1.5in}
\baselineskip 12pt
\centerline{{\large \bf ~~Measuring Hubble's Constant in our
Inhomogeneous Universe}}
\vspace{0.2in}
\centerline{~~XIANGDONG SHI\footnote{E-mail address:
shi@astro.queensu.ca},
LAWRENCE M. WIDROW\footnote{E-mail address:
widrow@astro.queensu.ca} and
L. JONATHAN DURSI\footnote{E-mail address: dursi@astro.queensu.ca}}
\vspace{0.2in}
\centerline{{\it ~~Department of Physics}}
\centerline{{\it ~~Queen's University, Kingston, Ontario, K7L 3N6, CANADA}}
\vspace{.5in}
\baselineskip=12pt
\centerline{ABSTRACT}
\bigskip
{ Recent observations of Cepheids in the Virgo cluster have bolstered
the evidence that supports a Hubble constant in $70-90\ksm$ range.
This evidence, by and large, probes the expansion of the Universe
within $100\mpc$.  We investigate the possibility that the expansion
rate within this region is systematically higher than the true
expansion rate due to the presence of a local, large underdense region
or void.  We begin by calculating the expected deviations between the
locally measured Hubble constant and the true Hubble constant for a
variety of models.  The calculations are done using linear
perturbation theory and are compared with results from N-body
simulations wherever possible.  We also discuss the expected
correlations between these deviations and mass fluctuation for the
sample volume.  We find that the fluctuations are small for the
standard cold dark matter as well as mixed dark matter models but can
be substantial in a number of interesting and viable nonstandard
scenarios.  However, deviations in the Hubble flow for a region of
radius $200\mpc$ are small for virtually all reasonable models.
Therefore, methods based on supernovae or the Sunyaev-Zel'dovich
effect, which can probe $200\mpc$ scales, will be essential in
determining the true Hubble constant.

We discuss, in detail, the fluctuations induced in the cosmic
background radiation by voids at the last scattering surface.  In
addition, we discuss the dipole and quadrupole fluctuations one would
expect if the void enclosing us is aspherical or if we lie off-center.
}
\baselineskip=12pt
\vspace{.5in}
\end{titlepage}
\newpage

\baselineskip=18pt

\vspace{24pt}

\section{Introduction}
\bigskip
It is straightforward to measure the Hubble constant in a perfectly
homogeneous and isotropic Universe: one simply divides the relative
velocity between any two points by their separation.  The situation is
of course far more difficult in our inhomogeneous and anisotropic
Universe.  For most cosmologists, the working assumption is that the
Universe becomes homogeneous on very large scales.  The goal then is
to find the recessional velocity and distance to an object
sufficiently far away so that the Hubble flow dominates over peculiar
velocities.  In this work, we address the possibility that
determinations of the Hubble constant have not yet reached such
distances.

Knowledge of the true, global value of the Hubble constant, $H_0$, is
central to any cosmological model.  Most measurements to date place
$H_0$ between $65-95\ksm$, values that are uncomfortably high given
estimates of the age of the Universe based on globular cluster dating,
and the theoretical prejudice that we live in a spatially flat,
matter-dominated universe.  In particular, if we accept that the
Universe is older than $12\,\rm Gyr$, then a Hubble constant $h>0.55$
($H_0=100\,h\ksm$) is incompatible with all $\om=1$ models ($\om$ is
the density of matter in units of the critical density) and $h>0.80$
is incompatible with all zero cosmological constant ($\Lambda=0$)
models.  In addition, a high value for the Hubble constant is, in
general, trouble for structure formation scenarios.  This is
especially true for cold dark matter (CDM) where the favoured value is
$h\sim 0.3-0.4$.  Finally, a low value for $H_0$ helps to close the
gap between the prediction for the baryon density from big bang
nucleosynthesis and recent determinations of the dark-to-baryonic
matter ratio in rich clusters (Bartlett et al. 1994).

All resolutions to the ``age problem'' come at a cost.  A true age for
the Universe significantly less than $12\gyr$ would require radical
rethinking of stellar evolution (Cage \& Walker 1992; Sandage 1993;
Chaboyer 1994; Shi 1995).  On the other hand, we can accommodate a
universe older than $12\gyr$ by requiring the Universe to be open, or
by invoking a non-zero cosmological constant.  For fixed $h$, an open
Universe corresponds to an older one; but this is problematic
especially if one accepts inflation which predicts $\Omega=1$ in
nearly all incarnations (see however Bucher et al.\,1995, Linde 1995).
Similarly, a universe dominated by a cosmological constant (or any
energy density with negative pressure) will be older than the
corresponding $\om=1$ universe.  The cost here is the introduction of
a new energy density term in the Friedmann equations whose origin is
completely mysterious.

In many respects, a true Hubble constant of $50\ksm$ or less
represents the simplest solution to the age problem.  However, the
majority of observations suggest that such values are ruled out.
These include recent measurements by the
Canada-France-Hawaii Telescope (CFHT) (Pierce et
al. 1994) and the Hubble Space Telescope (HST) (Freedman et al. 1994)
whose published values of
$h=0.87\pm 0.07$ and $h=0.80\pm 0.17$, respectively, exclude a
spatially flat, matter dominated universe.  Their
Cepheid results are also consistent with other determinations based on
techniques such as Tully-Fisher, planetary nebulae, and surface
brightness fluctuations (see van den Bergh 1992 for a review).  In
principle, Cepheids should provide the most accurate method for
determining the distance to Virgo.  At present, there is the potential
for large systematic errors (see, in particular, the paper by Freedman
et al. 1994) due primarily to uncertainties in the position of the
galaxy studied within Virgo.  Measurements of Cepheids in other
galaxies within the cluster will significantly reduce this uncertainty
and provide a distance determination to Virgo of better than 10\%.

Virgo is only $17\mpc$ away, far too close for a reliable measurement
of $H_0$.  But it is only one additional
rung of the distance ladder to clusters such
as Coma which, at a distance $\sim 100\mpc$, are assumed to be far
enough away that their recessional velocities are dominated by the
true Hubble flow.

But is this last assumption valid?  Or is it possible that the local
Hubble flow, as measured out to $\sim 100 \mpc$, overestimates the
true Hubble constant by 20-30\%?  The discovery of large voids and
sheet-like structures in redshift surveys (Kirshner et al. 1981; de
Lapparent et al. 1986; Geller \& Huchra 1989; da Costa 1991) and large
scale bulk flows (Lauer \& Postman 1994) on $\sim 100\mpc$ scales
should give us pause as variations in the Hubble flow are likely to
occur on comparable scales.

The question of deviations in the Hubble flow was addressed by Turner,
Cen, and Ostriker (1992, hereafter TCO) in the context of a simple
numerical experiment.  A model universe (with a specified value for
the global Hubble constant) is simulated using standard N-body methods
and it is assumed that
``observers'' in this universe can accurately measure the distances to
and radial velocities of {\em all} galaxies within a set distance $R$.  Each
observer constructs a Hubble diagram and reads off a value for the
local Hubble constant $\hl$.  In this way, the effects of local
deviations in the Hubble flow are separated out from
observer-dependent effects such as systematic errors in distance
determinations and incomplete sampling.  The distribution of $\hl$
measured by all observers in the simulation is studied and
compared with other statistics of the density and velocity fields such
as rms mass fluctuations and bulk flows.  In the end, one can make
probabilistic statements about the likelihood a given observer has for
measuring a local Hubble constant within the range allowed by
observations.

An alternate approach to this global statistical method is to explore
specific and simple models for our local region of the Universe.
Suppose, for example, we live in an underdense region or void.  Any
determination of the Hubble constant based on objects within this
region will reflect the local expansion rate rather than
the global one.  The simplest model of this type has us at
the center of a spherically symmetric section of an open spacetime
embedded in a spatially flat Friedmann-Robertson-Walker universe
(Suto et al.\,1994; Wu et al.\,1995; Moffatt \& Tatarski 1994).

In Section 2 we take a second look at the statistical approach
pioneered by TCO.  Our interest is in studying a variety of models and
so we use linear perturbation theory where probabilities can be
calculated directly from the linear power spectrum $P(k)$.  Our first
step is to compare linear theory results with those of TCO.  As
expected, agreement is good on large scales where perturbations are
small.  On small scales, linear theory tends to underestimate the rms
fluctuations in $\hl$.  In addition, TCO find that the distribution of
measured $\hl$ has a distinctly non-Gaussian shape.  We next use
linear theory to study correlations between $\hl$ and other
observables.  In particular, we derive the statistical relationship
between $\hl$ and the mass excess for the sample volume.

TCO consider the standard CDM scenario circa 1992 which is now known
to be inconsistent with results from the COBE DMR experiment
(Smoot et al.\,1992) and the APM galaxy survey
(Maddox et al.\,1990).  Here we use linear theory to calculate the rms
fluctuations in $\hl$ for a variety of observationally viable models.  As one
might expect, the size of deviations in the Hubble constant on
$100\mpc$ scales depends
sensitively on the shape of the primordial power spectrum but is
rather insensitive to the type of dark matter (i.e., hot, cold, or
mixed) in the model.

Section 3 deals with the more direct approach of modeling our local
region of the Universe.  The simplest scenario is to assume that we
are near the center of an underdense region that extends well past the
Coma cluster.  The mean density in this region must be fairly low
($\Omega_{\rm void}\la 0.5$) in order that there be a significant
difference between the true and measured values for the Hubble
constant.  This may not be so farfetched: there are hints
from the number counts of galaxies that we live in a very large
underdense region (Metcalfe et al.\, 1992) though
interpretation of this data is complicated by evolutionary effects
(see, for example, Loveday et al.\,1992).

A large underdense region today implies rather significant
perturbations in the energy density at the surface of last scattering.
These perturbations lead to fluctuations in the cosmic background
radiation (CBR)
through the Sachs-Wolfe effect.  We can therefore use CBR anisotropy
measurements to limit the number of voids
that reside in our Hubble volume.  In
addition, large angular-scale perturbations will be induced if our
local underdense region is aspherical, or if we lie off-center.  We
can therefore place important constraints on this scenario by using
the dipole and quadrupole anisotropy measurements.

Section 4 summarizes our results.  Our
general conclusion is that large fluctuations in $\hl$ for $R\simeq
100\mpc$ can occur, but only in nonstandard models.  For example, a
CDM model with a tilted primordial power spectrum $P(k)\propto k^n;
n>1.5$ will have $2\sigma$ fluctuations in $H$ in excess of $30\%$.
However, even in these nonstandard models, fluctuations in $H$ fall
rapidly with increasing scale: Determinations of $H$ based on objects
$\sim 200\mpc$ away should reflect the global value to within $10\%$
unless we live in an extremely rare, very large-scale underdense
region.

There is, at present, considerable discrepancy among measurements for
objects $\ga 200\mpc$ away.  The ``expanding photosphere method'' has
been used to determine distances to 18 type II supernovae as far away
as $180\mpc$ (Schmidt et al. 1994) and references therein).  An
analysis of this data finds $h=0.73\pm 0.06({\rm statistical})\pm
0.07({\rm systematic})\ksm$ with no evidence for spatial deviations in
the Hubble flow.  In addition Lauer \& Postman (1994) find that the
value of $H$ does not vary by more than 7\% between $30$ and $150
h^{-1}\mpc$.  On the other hand, measurements for very distant ($R\ga
400\hmpc$) objects based on the Sunyaev-Zel'dovich effect do seem to
yield systematically lower values for $h$ (Yamashita 1994; Birkinshaw
\& Hughes 1994) suggesting that we may in fact live in a low density,
high $h$ region of the Universe (Suto et al.\,1994).  The supernovae
results and Lauer \& Postman data seem to indicate that this void
would be much larger than anything expected, even for fairly
nonstandard models designed to give more power on large scales.

\section{Hubble Constant Statistics}

\subsection{Linear Theory Analysis}

In linear perturbation theory (Peebles 1993), the relationship between the
peculiar velocity field and the density contrast
$\delta({\bf x})\equiv (\rho({\bf x})-\rho_b)/\rho_b$ is given by
\be{velocity}
{\bf v}({\bf x})~=~\frac{2}{3H_0}\int d^3y\,\frac{{\bf x-y}}{|{\bf
x-y}|^3}\,\delta({\bf y})
\ee
where $\rho({\bf x})$, $\rho_b$ and $H_0$ are the density field,
background density, and global Hubble constant today.  The peculiar
Hubble flow, $\delta H({\bf x})\equiv H_L({\bf x})-H_0$ is
\be{localh}
\delta H({\bf x})~=~
\int d^3 y \,
{\bf v}({\bf y})\cdot\frac{ {\bf x}-{\bf y}}
{|{\bf x}-{\bf y}|^2}W({\bf x}-{\bf y})
\ee
where $H_L({\bf x})$ is the local value of the Hubble constant as
measured by an observer at position ${\bf x}$ and $W({\bf x}-{\bf y})$
is the window function for the observations.  $\delta H$ is
essentially a breathing mode (expansion or contraction) of the
peculiar velocity field, corresponding to the trace of the shear
tensor (see, e.g. Kaiser 1988).  For a perfectly volume limited sample
out to a distance $R$, $W$ is a step function, $W({\bf x}-{\bf
y})=\theta(R-|{\bf x}-{\bf y}|)V_W^{-1}$ where $V_W\equiv 4\pi R^3/3$.
In a Friedmann-Robertson-Walker universe, $\delta H\to 0$ for
$R\to\infty$.

Equations \EC{velocity} and \EC{localh} can be written in terms of the
Fourier transform of the density contrast:
\be{ftlocalh}
\frac{\delta H}{H_0}~=~
\int \frac{d^3k}{(2\pi)^{3/2}}\,\delta({\bf k})\,{\cal Z}(kR)\,e^{i{\bf k\cdot
x}}
\ee
where $\delta({\bf k})\equiv \left (2\pi\right )^{-3/2}
\int d^3x\delta ({\bf x})e^{i{\bf k\cdot x}}$,
\be{window}
{\cal Z}(x)~=~3\,\frac{ \sin(x) - {\rm Si}(x)}{x^3}\, ,
\ee
and ${\rm Si}(x)\equiv \int_0^x dx \sin(x)/x$.
The rms fluctuations in the local Hubble constant $\delta_H\equiv
\langle\left (\delta H/H_0 \right )^2\rangle^{1/2}$
($\langle\dots\rangle$ denotes spatial average) is given by
\be{rmslocalh}
\delta_H^2~=~\frac{1}{2\pi^2}\int k^2 dk P(k){\cal Z}^2(kR)
\ee
where $P(k)=|\delta ({\rm k})|^2$ is the power spectrum.  If the
primordial perturbations are Gaussian, the distribution of $\delta_H$
will be Gaussian, at least in the linear regime.

\subsection{TCO and Linear Theory}

TCO use numerical simulations of cosmological models to generate a set of
hypothetical observers.  These observers can accurately measure the distances
to and velocities of the galaxies within a pre-set distance.
The local value of the Hubble constant, as measured by the $k$th
observer, is given by
\be{tcolocalh}
H_k~=~\frac{1}{N}\sum_{i\ne k}\frac{{\bf r\cdot v_i}}{|{\bf
r}|^2} ~+~H_0
\ee
where ${\bf r_i}$ and ${\bf v_i}$ are the position
and velocity vectors of the $i$th galaxy, ${\bf
r}\equiv {\bf r_i-r_k}$, and the sum is over all $N$ galaxies in the
prescribed volume.

Eq \EC{tcolocalh} is essentially the discrete version of Eq
\EC{localh} with the important distinction that Eq \EC{localh}
involves a volume average whereas Eq \EC{tcolocalh} involves a sum
over galaxies.  Since galaxies tend to reside in high density regions
(this will depend on biasing, i.e., galaxy formation) where the
expansion rate is slower, the results of TCO will tend, on average, to
be skewed towards negative values of $\delta H\equiv\left
(H_k-H_0\right )/H_0$.  This is indeed found to be the case.

Column 1 of Table 1 summarizes the results for $\delta_H$ from TCO.
The results derived from linear theory for their model
(baryon density, in units of the critical density, $\Omega_B=0.05$;
$h=0.5; \, \om+\Omega_B=1$; rms mass fluctuation on $8 h^{-1} \mpc$,
$\sigma_8=0.67$)
are shown in Column 2.  As expected, the agreement is best
on large scales where perturbations are small.  On small scales,
linear theory tends to underestimate $\delta_H$.  In any case, it is
apparent that for this particular model, the local value of the Hubble
constant, as measured by observers with complete coverage out to
$100\mpc$, is expected to be with 5\% or so of the true or global
value.

\begin{table}
\begin{center}
\begin{tabular}{|| c | c | c | c | c ||}\hline
{}~R (Mpc)~~ &  ~~$\delta_H$ (TCO)~~& ~~~ $\delta_H$ (LT)~~~ & ~~~ $m$
(LT) ~~~ & ~~~ $\sigma_{HM}$ ~~~ \\ \hline
10    & 1.08 &  0.42 & -0.64 & 1.07 \\ \hline
20    & 0.45 &  0.23 & -0.64 & 0.79 \\ \hline
40    & 0.18 &  0.11 & -0.65 & 0.55 \\ \hline
60    & 0.10 &  0.071 & -0.67 & 0.43 \\ \hline
80    & 0.062 & 0.049 & -0.67 & 0.35 \\ \hline
120   & 0.029 & 0.027 & -0.67 & 0.26 \\ \hline
\end{tabular}
\end{center}
\caption{TCO and Linear theory results for $\delta_H$}
\end{table}

\subsection{Correlations between $\delta_H$ and other observables}

A volume limited measurement of $H$ cannot separate the peculiar
Hubble flow within that volume from the true Hubble flow.  There are,
however, other quantities accessible to observers which characterize
the peculiar velocity field and matter distribution.  Correlations
between such quantities and the peculiar Hubble flow, if they exist,
would provide a means for correcting our measurements of $\hl$.

An obvious quantity to consider is mass fluctuation within the
sample volume, $\delta M/M$.  For an observer at the center of a
spherically symmetric region with constant density contrast, $\delta
H/H_0=-\frac{1}{3}\delta M/M$ so long as $\delta \ll 1$.  If this were
the case, knowledge of $\delta M/M$ would be sufficient to correct
local measurements of $H$.  In general, however, there is only a
statistical relationship between $\delta H/H_0$ and $\delta M/M$.  TCO
explore this relationship by plotting $\delta M/M$ versus $\delta
H/H_0$ for the observers in their simulation (their Figure 6a).  A
linear fit is obtained by minimizing the total orthogonal distance of
all points on this figure to the straight line
\be{fit}
\frac{\delta H}{H_0}~=~m\frac{\delta M}{M}~+~b\, ;
\ee
that is, by minimizing the function
\be{chisq}
\chi^2~=~\frac{1}{1+m^2}\sum_i
\left (\frac{\delta H_i}{H_0}-m\frac{\delta M_i}{M}-b\right)^2\,.
\ee
Their result,
$m=-0.60, ~b=-0.008$, corresponds to a slope that is nearly twice what
one would naively expect from linear theory,
and they attribute this difference to nonlinear effects.
While they find substantial scatter in the plot, they conclude that
$\delta M$ offers the best hope for correcting local measurements of
the Hubble constant.

Interestingly enough, the results of TCO are similar to those found in
linear theory.  Consider the linear theory expression for the
fractional mass excess in a spherical top-hat region:
\be{deltamass}
\frac{\delta M}{M}~=~
\int \frac{d^3k}{(2\pi)^{3/2}}\,\delta({\bf k})\,W(kR)\,e^{i{\bf k\cdot
x}}
\ee
where
\be{window2}
W(x)~=~3\,\frac{ \sin(x) - x\cos(x)}{x^3}~.
\ee
The rms mass fluctuation is given by
\be{rmsmass}
\delta_M^2~=~\frac{1}{2\pi^2}\int k^2dkP(k)W^2(kR).
\ee
In the limit $kR\to 0$, ${\cal Z}\to -1/3$ and $W\to 1$ which would
seem to support the naive expectation.  However, the window functions
in Eq \EC{rmslocalh} and \EC{rmsmass} differ not only in amplitude but
in shape with ${\cal Z}$ having more support at larger $k$.  A linear
fit, analogous to the one calculated in TCO, yields
\be{slope}
m = \frac{
\delta_H^2 - \delta_M^2
+ \sqrt {
\left ( \delta_H^2 - \delta_M^2\right )^2  +
4 \left < \frac{\delta M}{M} \frac{\delta H}{H} \right >^2 }
} { 2 \left < \frac{\delta M}{M} \frac{\delta H}{H} \right > }
\ee
where
\be{crosscorr}
\left < \frac{\delta M}{M} \frac{\delta H}{H}\right >
{}~=~\frac{1}{2\pi^2}\int k^2dk P(k){\cal Z}(kR)W(kR)
\ee
The rms fluctuation from the best fit curve $\sigma_{HM}$ is given
by
\be{sigma}
\sigma_{HM}^2 =
\frac{\delta_H^2  + m^2 \delta_M^2 -
2 m \left < \frac {\delta M}{M} \frac{\delta H}{H} \right >}
{ \left ( 1 + m^2 \right ) }.
\ee
Our results, shown in Table 1, illustrate that those of TCO can be
largely explained by linear theory.

In principle, one can use knowledge of $\delta M/M$ in our region of
the Universe to correct local measurements of $H$.  With this in mind,
we choose to determine the correlation between $\delta H/H_0$ and
$\delta M/M$ by a slightly different procedure.  Specifically, we fit
minimize the distance of $\delta H/H_0$
to the curve given by Eq \EC{fit}, i.e., we minimize
\be{chisq1}
\chi^2~=~\sum_i
\left (\frac{\delta H_i}{H_0}-m\frac{\delta M_i}{M}-b\right)^2\,.
\ee
A standard least squares analysis gives $m=\left <
\frac{\delta M}{M} \frac{\delta H}{H}\right >/\delta_M^2$
and $\sigma_{HM}^2=\delta_H^2-2m\left <
\frac{\delta M}{M} \frac{\delta H}{H}\right >+m^2\delta_M^2$.
We use these formulae to calculate $m$ and $\sigma_{HM}$ for the various
models discussed above and give the results in Table 2.

\subsection{Standard CDM and Variations}

We calculate $\delta_H$, $m$, and $\sigma_{HM}$ for a variety of
models specified by their power spectra $P(k)$.  We begin by
considering a family of models where the initial power spectrum is an
$n=1$, Harrison-Zel'dovich power law: $P_{\rm i}(k)=Ak$.  The power
spectrum today is given by $P_i$ times an appropriate transfer
function: $P(k)=AkT^2(k)$.  Following Efstathiou, Bond, \& White (1992)
we choose the following parametric form for $T(k)$:
\be{cdmps}
T(k)=\left (1+\left (ak/\Gamma +\left (bk/\Gamma\right )^{3/2}+\left
(ck/\Gamma\right )^2\right )^\nu\right )^{-1/\nu}
\ee
where $a=6.4h^{-1}\mpc$, $b=3.0h^{-1}\mpc$, $c=1.7h^{-1}\mpc$, and
$\nu=1.13$.  The power spectrum is therefore specified by $\Gamma$ which
determines the shape and $A$, or equivalently $\sigma_8$ which sets
the normalization.  $\delta_H\propto\sigma_8$ and so, in
Figure 1, we show $\delta_H$ as a function of $\Gamma/\sigma_8$.  The
results are given for $h=0.5$
and three different values of $R$.  As expected, $\delta_H$ increases
with decreasing $\Gamma$.

The Lauer \& Postman (1994) velocity data on Abell clusters shows an
large scale bulk motion that may indicate power on very large scales
in excess of what is expected in most COBE normalized models.  While
there is still some controversy over the interpretation of this data
(Strauss et al 1994, Feldman \& Watkins
1994) it is useful to consider the implications these results would
have for determinations of the Hubble constant, the idea being that a
universe with large scale bulk flows might also have large breathing
mode fluctuations in the peculiar velocity field that foil our
attempts to measure $H$.  Jaffe \& Kaiser (1995) apply a likelihood
analysis to the Lauer \& Postman (1994) data using Eq \EC{cdmps}, and
find a peak in the
likelihood function at $\sigma_8\simeq 0.3$ and $\Gamma\simeq 0.025$.
{}From Figure 1, we see that this implies a rather modest $\delta_H$:
$\delta_H(R=100\mpc)=0.05$.  They note however, that there is a rather
wide range of parameters that are consistent with the Lauer \& Postman
data.  We see from their Figure 1, for example, that $\Gamma=0.35$ and
$\sigma_8=1.2$ is within the 68\% confidence intervals for the COBE
quadrupole measurement and also within the 50\% likelihood contour for the
Lauer
\& Postman data.  For these parameters, $\delta_H\simeq 0.12$.

We next consider the following models: (1) Standard CDM (sCDM) with
COBE normalization (Bunn, Scott, \& White 1994): This is essentially
the model considered in Section 2.2 but the $\sigma_8=1.38$.  (2)
Mixed dark matter (MDM): We use Holtzman's (1989) linear power
spectrum for a 70\% cold, 30\% hot mixed dark matter model.
Normalization is the same as in Model 1.  (3) Tilted primordial $P(k)$
with spectral index $n>1$.  Models with both CDM and MDM are used.
Again we normalize to the COBE results on large scales following Bunn,
Scott, \& White (1994).  The results for these models are summarized
in Table 2.

\begin{table}[ht]
\begin{center}
\begin{tabular}{|| r | c | c | c | c | c ||}
\hline & & & \multicolumn{2}{c}{ Tilted CDM } & MDM     \\

 &  sCDM  & MDM & n=1.15 & n=1.5 & n=1.5  \\ \hline

$ \delta_H ( 100~ {\rm Mpc} )$ & 0.074 & 0.072 & 0.098 & 0.17  & 0.16  \\
$ \delta_H ( 150~ {\rm Mpc} )$ & 0.040 & 0.041 & 0.052 & 0.083 & 0.083 \\
$ \delta_H ( 200~ {\rm Mpc} )$ & 0.025 & 0.026 & 0.032 & 0.049 & 0.050 \\
\hline
$ (\delta H/H) $ vs. $ (\delta M/M)$
			       &       &       &       &       &       \\
$m (150~ {\rm Mpc})$           &--0.60 &--0.62 &--0.60 &--0.57 &--0.59 \\
$\sigma (150 ~{\rm Mpc})  $    &  0.10 &  0.10 &  0.13 &  0.18 &  0.10 \\
\hline
\end{tabular}
\end{center}
\caption{Linear-theory $\delta_H (R)$, $m$, and $\sigma$ for different
power spectra $P(k)$.}
\end{table}
Evidently, the rms peculiar Hubble flow on $100\mpc$ scales can be
quite substantial, especially for $n>1$.  These so-called
blue primordial spectra (Lucchin et al.\,1995) were introduced to
boost power on $100\mpc$ scales relative to COBE scales.  Indeed, it
is worth noting that the constraints on $n$ from the COBE DMR
experiment are not very restrictive, and $n=1.5$ is easily allowed by
the data (Bennett et al.\,1994).  A CDM model with $n>1$ will no
doubt lead to excessive power on small ($\la 1\mpc$) scales and this
has motivated Lucchin et al.\,(1995) to consider $n>1$ with MDM.
Another potential difficulty with these models is that they can lead
to large temperature fluctuations in degree-scale CBR experiments.
Consider the usual decomposition of the CBR anisotropy:
$\delta T/T(\theta,\phi)=\sum_{lm}a_{lm}Y_{lm}(\theta,\phi)$.  For
Gaussian perturbations, the temperature fluctuations are completely
specified by the variance $C_l\equiv\langle |a_{lm}|^2\rangle$.
Following Kosowsky \& Turner (1995) we estimate the effects of tilting
the spectrum to be
\be{tilt}
C_l(n)\simeq \frac{l}{2}^{\ln(l/2)\left (n-1\right )}C_l(n=1)
\ee
For $l=100$ we find that the temperature fluctuations ($\propto
C_l^{1/2}$) are increased by a factor of $3$ over the $n=1$ model for
$n=1.15$ and $50$ for $n=1.5$.  The latter case is clearly
unacceptable unless degree-scale fluctuations are damped by a period
of early reionization.  In any case, the next generation of CBR
experiments will lead to tighter constraints on the primordial
spectrum and help settle this issue.

\section{CBR Fluctuations from Voids}

We have seen that the possibility of living in a large underdense
region opens up the prospect for solving the $H_0/t_0$ problem without
giving up spatial flatness or resorting to a cosmological constant
(Suto et al. 1994; Wu et al. 1995; Moffatt \& Tatarski 1994).  In the
simplest scenario, we would be at the center of an underdense region
containing all of the objects used to measure $H$.  We could then
neglect safely the dynamics of the wall that separates this region from the
rest of the Universe.  For the CFHT and HST results this region would
include the Coma cluster as well as clusters used to determine the
peculiar velocity of Coma (as in, for example, Han \& Mould (1992)).
The region would therefore have to have a radius {\it significantly larger
than} $100\mpc$.  An even larger region would be required if one is to
accommodate results from type II supernovae (e.g., Schmidt et
al.\,(1994) which draw from objects up to $180\mpc$ away.  On the
other hand, it may be possible to accommodate the CFHT and HST results
with a local underdense region whose radius is {\it comparable to}
$100\mpc$.  This could arise if the Local Group
and/or clusters such as Coma could lie on the surface of a void or
perhaps on a ridge separating merging voids.  Using N-body simulations
to study interacting voids, Dubinski et al.\,(1993) find significant
peculiar velocities tangential to the walls that separate merging
voids, peculiar velocities that might help explain an anomalously high
measurement of $H$.  However, this scenario is inherently more
complicated that the scenario in which the void contains the clusters
used to determine $H$.  In particular, the large scale peculiar
velocity field is no longer characterized by a simple breathing mode.

The mean density in our region of the Universe must be fairly small if
local measurements of $H$ are to be significantly higher than the
global value.  In Figure 2 we show $H_L$, the expansion rate inside a
uniform region, as a function of the density in that region, $\ov$.
For $|\ov-1|\ll 1$, we have the linear result discussed above:
$H_L-H_0~=~\frac{1}{3}H_0\left (1-\ov\right )$.  The HST and CFHT
results with quoted error bars are also shown.  We see that
$t_0>12\gyr$ and $\hl>63\ksm$ (the $1\sigma$ lower bound for the HST
result) imply a local density $\ov\la 0.55$,
i.e., a nonlinear void.

Voids somewhat smaller than the ones required here appear to be quite
common.  Redshift surveys reveal a network of voids with typical
diameters of $50-60\mpc$ (Kirshner et al. 1981; de Lapparent et
al. 1986; Geller \& Huchra 1989; da Costa 1991).  There is also
evidence for larger voids (Einasto, Joeveen, \& Saar 1980, Bahcall \&
Soneira 1982) though, as one might expect, the larger voids are not
nearly so empty.  Of course, redshift surveys give the distribution of
visible galaxies and not the underlying matter distribution and so it
is difficult to say what the true density contrast is inside the
voids.  Nevertheless, a picture emerges of a hierarchy of voids in
which the density contrast decreases with increasing scale.

Galaxy counts and redshift surveys offer the possibility of directly
probing the density and velocity fields in our region of the Universe.
However, they are each plagued by a number of difficulties.  In
particular, evolutionary effects can mask or mimic density
inhomogeneities in the galaxy counts.  In addition, there is the
question of how one normalizes this data.  On the other hand, attempts
to map the peculiar velocity field must contend with sparse sampling;
the rich clusters used to probe the velocity field are just too rare.

Observations of anisotropies in the CBR present an alternative, and
potentially clean probe of large scale density inhomogeneities.  In
this section we focus on constraints of large scale voids from CBR
experiments.  Previous attempts along these lines have used a
statistical approach (Blumenthal et al. 1992; Piran et al. 1993).  The
idea was to construct a power spectrum that leads to a ``reasonable
number'' of voids with the desired size and density contrast and to
then see if this power spectrum is consistent with CBR observations.
The discussion at the end of this section follows this
approach.  Here we focus on temperature fluctuations from individual
voids and the primordial fluctuations that gave rise to them.  In
particular, we consider three distinct sources of temperature
fluctuations: (1) primeval density fluctuations at the last scattering
surface (primeval voids); (2) evolving voids between us and the last
scattering surface (intermediate voids); and (3) the void enclosing
us.

\subsection{Formalism}

Consider a spherically symmetric density perturbation in a matter
dominated universe.  The metric can be written (e.g., Kramer et
al. 1980)
\begin{equation}
ds^2=-dt^2+e^{\lambda (\chi,t)}d\chi^2+r^2(\chi,t)d\Omega^2~.
\end{equation}
It is convenient to define the (spatially dependent) scale factor
$a(\chi,t)$:
\begin{equation}
r(\chi,t)=a(\chi,t)\chi~ .
\end{equation}
Let $a_i(\chi)=a(\chi,t_i)$ and $\rho_{{\rm b}i}$ be the scale factor
and background density at some initial time $t_i$ and let
$\dbc$ be the initial average density contrast enclosed by the
$\chi=constant$ surface:
\be{deltachi}
\dbc=\bar\rho_i(\chi)/\rho_{{\rm b}i}-1.
\ee
where $\bar\rho_i(\chi)$ is the average density contrast enclosed by
the $\chi=constant$ surface.
The equation of motion for $a(\chi,t)$ follows from the Einstein equations:
$$
{a(\chi,t)\over a_i(\chi)}~=~{1+\dbc\over 2\dbc}
{{\rm d}f(\eta)\over{\rm d}\eta}
\quad\quad\quad\quad\quad\quad\quad\quad\quad
$$
\be{eofm}
H_i(\chi)[t+t_{\rm c}(\chi)]~=~
{1+\dbc\over 2\vert\dbc\vert^{3/2}}
{\dbc\over\vert\dbc\vert}f(\eta)
\ee
where $f(\eta)=\eta-\sin[\rm h]\eta$ for $\delta>[<]0$ and
$H_i(\chi)=(\dot a/a)\vert_{t=t_i}$.

For a photon passing through the origin
\be{dchidt}
{{\rm d}\chi\over {\rm d}t}=\pm {\sqrt{1-(4/9)\delta(\chi)\chi^2/t_i^2}
\over r^\prime}\, ,
\ee
(Fang and Wu 1993) where the plus (minus) sign is for a photon moving
away from (towards) the origin.  The frequency shift of the photon is
\be{shift}
{\nu_i\over \nu_f}={\rm exp}\left(\int^{t_f}_{t_i}{1\over 2}\dot\lambda dt
\right)={\rm exp}\left (\int^{t_f}_{t_i} {\chi\dot a^\prime+\dot a\over
\chi a^\prime +a}dt\right ),
\end{equation}
where $\nu_i$ is the frequency of the photon emitted at $t_i$ and
$\nu_f$ is the frequency of the photon observed at $t_f$.  Here and
throughout, we use overdot to denote partial derivative with respect
to $t$ and prime to denote partial derivative with respect to $\chi$.
The resultant temperature fluctuation for a blackbody distribution of
photons passing through the center of the density fluctuation, in
a spatially flat, matter-dominated universe, is
\be{deltat}
{\delta T\over T}={\nu_f^{\rm fluc.}-\nu_f^{\rm bkgd}\over\nu_f^{\rm bkgd}}=
\left({t_i\over t_f}\right )^{2/3}
{\rm exp}\left(\int^{t_f}_{t_i} {\chi\dot a^\prime+\dot a\over
\chi a^\prime +a}dt\right )-1\, .
\ee
By our convention, a negative $\delta T/T$ implies a cold spot in CBR
sky.

The frequency shift of a photon passing through the center of a void
is found by numerically integrating Eq \EC{deltat}.  The calculation
is straightforward when the density perturbation is linear but requires
a special approximation, to be discussed in section 3.3, once the
void becomes nonlinear. Spherical symmetry is assumed for the voids
through out our calculation, unless explicitly stated otherwise.

\subsection{Primeval Voids at the Last Scattering Surface}

We assume that the voids present today result from the gravitational
amplification of primordial perturbations.  The alternative would be
to have the voids produced after recombination by some
non-gravitational process.  The evolution of the spacetime inside a
void is calculated from Eq \EC{eofm}.  For definiteness, we assume a
density profile of the form:
\begin{equation}
\dbc=\left\{\matrix{\dbi,&\quad\chi<\chi_1.\cr
-\dbi\chi_1^3/(\chi_0^3-\chi_1^3),&\quad\chi_1\le\chi<\chi_0.\cr
0,&\quad\chi\ge\chi_0.}\right \}
\label{profile}
\end{equation}
so that the ratio $\chi_1:\chi_0$ determines whether one has a thick
or thin wall.
In Figure 3, we plot $\dbi$ as a function of $\ov$ today.  We see,
for example, that $\ov\la 0.55$ requires $\dbi\la -8.5\times
10^{-4}$ at recombination.

The correspondence between the size of the primeval void and its size
today is trickier. In the linear regime, the size of the void can be
easily calculated from Eq \EC{eofm}.  In the nonlinear regime, matter
from both inside and outside the void piles up on its surface (Maeda
\& Sato 1983a; Berschinger 1985).  In any case, the size of the void
can grow no faster than the expansion rate inside the void:  A void
that has $\Omega_{\rm void}=0.55$ today grew no more than 20\% (in
comoving coordinates) from the time of recombination.

Density perturbations at the time of recombination lead to temperature
fluctuations in the CBR sky.  If the scale of the perturbation $l$ is
much larger than the horizon size at recombination, $ct_{\rm rec}$,
the temperature fluctuations will be dominated by the Sachs-Wolfe
effect (Sachs \& Wolfe 1968).  The expected temperature fluctuation is
\be{sweffect}
{\delta T\over T}\sim -{\delta\rho\over\rho}
\left({l\over ct_{\rm rec}}\right)^2.
\ee
where we have omitted a prefactor of order $0.1$.
For $l\la ct_{\rm rec}$, other effects such as the Doppler peak and
adiabatic fluctuations in baryons become important (Hu, Sugiyama, \&
Silk 1995).

Figure 4 shows our numerical calculation of the Sachs-Wolfe effect for
a primeval void with $\dbi=-8.5\times 10^{-4}$ and
$\chi_1=67h^{-1}/(1+z_{\rm rec})$ Mpc and three different density
profiles.  $\chi_{\rm LSS}$ is the distance from the last scattering
surface to the center of the void with negative values corresponding
to the situation in which the last scattering surface lies behind the
center of the void.  In these calculations, we neglect the thickness
of the last scattering surface since it is unimportant so long as
$\chi_1\gg 10h^{-1}/(1+z_{\rm rec})$ Mpc.  Note that the temperature
fluctuations scale linearly with $\dbi$.

We see that sign and magnitude of the temperature fluctuation depends
sensitively on where the last scattering surface intercepts the void.
In addition, while the maximum temperature fluctuation is relatively
insensitive to the thickness of the wall, the range in values of
$\chi_{\rm LSS}$ over which $\delta T/T$ is large increases as we
increase the thickness of the wall.

We wish to calculate the number of CBR fluctuations as a function of
angular diameter and amplitude.
Let $n^{\rm rec}_{\rm void}$ be the number density of voids that lead
to CBR fluctuations with the desired amplitude and angular size.
The expected number of fluctuations seen by a given experiment will be
\begin{eqnarray}
\langle{\cal N}\rangle & \simeq & 4\pi\left({3ct_0\over 1+z_{\rm rec}}\right)^2
n^{\rm rec}_{\rm void}\chi_1{\cal P} \nonumber \\
& \approx & 14h^{-3}{\rm \ Mpc}^3 \, \cdot \,
n^{\rm rec}_{\rm void}
\left[{\chi_1\over 100\mpc/(1+z_{\rm rec})}\right]{\cal P}
\end{eqnarray}

The COBE DMR experiment has mapped the entire sky with $7^\circ$
resolution.  Primeval voids with $\chi_1\ga 350h^{-1}/(1+z_{\rm rec})$ and
$\dbi\la -8.5\times 10^{-4}$ lead to fluctuations that would have
been easily seen in this experiment.  We therefore require
$\langle{\cal N}\rangle\la 1$ for these very large voids.  This
implies a tight constraint on their number density today,
$n^0_{\rm void}=n^{\rm rec}_{\rm void}/(1+z_{\rm rec})^3$:
\begin{eqnarray}
n^0_{\rm void} & \la  & 2\times 10^{-11}h^3\left[{100/(1+z_{\rm rec}){\rm
Mpc}\over
\chi_1}\right]{\rm \ Mpc}^{-3} \nonumber \\
& \la & 5\times 10^{-11} ~\mpc^{-3}.
\end{eqnarray}

Degree-scale measurements of temperature fluctuations in the CBR place
constraints on smaller-scale voids.  The MAX and MSAM experiments
cover $\sim 40$ square degrees in total (${\cal P}\simeq 0.001$) with
$0.5^\circ$ resolution and
find an rms temperature fluctuation of $\vert\delta T/T\vert\la
6\times 10^{-5}$ (Meinhold and Lubin 1991; Alsop et al. 1992; Meinhold
et al. 1993; Gundersen et al. 1993; Devlin et al.  1994; Clapp et
al. 1994; Cheng et al. 1994).  Figure 5 shows the absolute value
of $\delta T/T$ in the direction of the center of the void, averaged over
$-0.5<\chi_{\rm LSS}/\chi_0<0.5$. From Figure 5, we see that primeval
voids with $\dbi\le -8.5\times 10^{-4}$,
$\chi_1\ga 50h^{-1}(1+z_{\rm rec})^{-1}$ Mpc, and
$\chi_0:\chi_1=3:1$ (thick wall case) or
$\chi_1\ga 80h^{-1}(1+z_{\rm rec})^{-1}$ Mpc
and $\chi_0:\chi_1=9:8$ (thin wall case) can easily lead to temperature
fluctuations that exceed the observational limit.
Several individual patches with $\vert\delta T/T\vert$ as large as
$1\times 10^{-4}$ have been observed though it is not as yet
clear if these rare large fluctuations represent true CBR anisotropies
or foreground contamination.
In any case, primeval voids with $\dbi\le -8.5\times 10^{-4}$
and $\chi_1\ga 50h^{-1}(1+z_{\rm rec})^{-1}$ Mpc (thick wall)
or $\chi_1\ga 80h^{-1}(1+z_{\rm rec})^{-1}$ Mpc (thin wall) should be rare
in the currently covered area of MAX and MSAM.
To be conservative, we require $\langle{\cal N}\rangle<10$.  This
yields the constraints
\begin{equation}
n^0_{\rm void}\la 3\times 10^{-7}h^3 {\rm Mpc}^{-3}.
\end{equation}
This implies that $\la {\cal O}(0.1)$
of the volume of our universe can be occupied by these voids.

\subsection{Voids at Intermediate Redshift}

Temperature fluctuations due to nonlinear structures between us and
the last scattering surface were first considered by Rees and Sciama
(1968).  For subhorizon-sized structures, the temperature fluctuation
will be ${\cal O}((R/ct)^3)$ where $R$ is the physical size of the object at
the time $t$ when the photon crosses it.

We use the formalism of Section 3.1 to calculate temperature
fluctuations due to an intermediate nonlinear void of arbitrary
interior density.  For simplicity, we assume the a thin wall separates
the void from the rest of the Universe.  For the special case of a
vacuum void, Thompson and Vishniac (1987), and Mart\'inez-Gonz\'alez
et al. (1990), find
\be{vacvoid}
{\delta T\over T}\approx\left({R\over ct}\right)^3
\cos\Psi\left({8\over 9}\gamma-{16\over 27}-{16\over 81}\cos^2\Psi\right)
\ee
where $\Psi$ is the angle between the photon's direction and a line
from the center of the void to the exit point of the photon.
$\gamma$ describes the expansion of the void:
$R\propto t^\gamma$.

The numerical integration of Eq \EC{deltat} must be modified in order
to handle the nonlinear voids considered here.  In the formalism
outlined above, the initial density perturbation is divided into
concentric, spherical shells.  However, once the density fluctuation
becomes nonlinear and matter starts to pile up on the wall of the
void, shell-crossing will occur.  To avoid complications due to
shell-crossing, we choose instead to divide the density profile into
concentric shells at the time when the photons exit the void.  The
``evolution'' of the void is now found by solving the same equations
as before, but letting time run backwards.  However now, there is no
shell crossing and the frequency shift of the photons can be
calculated using the same methods as in section 3.1.

The expansion rate inside the void is still calculated by evolving the
initial perturbation (i.e., the perturbation at recombination)
forward in time.  This calculation not only determines $H_L$ but
also the velocity perturbation associated with the void at the time
when the photons exit the void.  The latter is required for our
backward-time integration.  The index $\gamma$ for the expansion of
the wall is put in by hand.  In general, $0.667\le\gamma\le 0.8$.
$\gamma=2/3$ corresponds to the case where the wall expands with the
background, a situation that will arise when the wall collides with
the walls enclosing neighboring voids.  $\gamma=0.8$ corresponds to a
single isolated void.  The result is easily derived from either energy
or momentum conservation (Maeda and Sato 1983a; Berschinger 1985).

For a nearby vacuum void with a thin wall and a radius
of 100$h^{-1}$ Mpc, eq. \EC{vacvoid} yields approximately $-1\times
10^{-5}$ for $\gamma=0.8$, and $-2.5\times 10^{-5}$ for $\gamma=2/3$
in the direction of the center of the void.  Our calculations yield
$-7\times 10^{-6}$ and $-2.6\times 10^{-5}$ respectively. Our
calculations also confirmed the scaling law of $\delta T/T\propto
(R/ct)^3$.  Empirical comparisons indicate that our results are
reliable to order ${\cal O}(10^{-6})$.

Figure 6 shows our results for a void with a 100$h^{-1}$ Mpc radius
and with different $\Omega_{\rm void}$ under a thin wall
approximation.  $\delta T/T$ is the CBR temperature fluctuation in
the direction through the center of the void. It is assumed that we
are located just outside the wall. The figure indicates that the
$\delta T/T$ due to a nearby 100$h^{-1}$ Mpc void is not large enough
to be constrained by COBE measurements. However, according to the
$(R/ct)^3$ scaling law, voids with a radius $\ga 200h^{-1}$ Mpc (thus
yielding $\ga 8$ times more temperature fluctuation) and a
sufficiently low $\Omega_{\rm void}$ can be in conflict with the COBE
observation.

\subsection{The Void Enclosing Us}

For spherical voids between us and the last scattering surface, the leading
order (${\cal O}(R/ct_0)$) and the second order (${\cal O}(R/ct_0)^2$)
contributions to the CBR anisotropy cancel (Rees
\& Sciama 1968).  For the void enclosing us, this cancellation does not
occur and we have a contribution to the dipole component of the CBR of
${\cal O}(R/ct_0)$.  This contribution can be viewed simply as arising from
our peculiar motion relative to the CBR rest frame and its amplitude
is easy to estimate:
\begin{equation}
\left\vert{\delta T\over T}
\right\vert_{\rm dipole}=\left(H_{\rm void}t_0-
{2\over 3}\right){r\over ct_0},
\label{dipole}
\end{equation}
Here $r$ is the our distance to the center of the void, as illustrated
in the geometry of Figure 7.  A detailed calculation, making use of
the formalism developed above, leads to similar results.  In Figure 8,
we show the amplitude of the dipole moment as a function of $r$ for
$0\le\Omega_{\rm void}\le 0.55$.  The requirement that $|\delta
T/T|\la 10^{-3}$ leads to the constraints $r\la 7\hmpc$ for a vacuum
void, and $r\la 22\hmpc$ for $\ov=0.55$.

The contribution to higher order moments from the void enclosing us
is ${\cal O}((R/ct_0)^3)$ and can lead to unacceptably large fluctuations for
$R\ga 200h^{-1}$ Mpc, unless we are essentially at the center.

Thus far we have assumed perfect spherical symmetry for the voids.
This is no doubt unrealistic, though it does appear that voids become
more spherical with time (Blaes et al. 1990).  Clearly, deviations
from spherical symmetry will lead to CBR distortions.  Consider, for
example, a bump in the wall with height $\Delta R$ (Figure 7).  We
estimate of its order of magnitude as follows.  The deviation from
spherical symmetry should contribute to temperature fluctuation at
orders $(R/ct_0)^m(\Delta R/R)^n$, where $m$ and $n$ are positive
integers. The leading order, with $m=n=1$, is simply a dipole
resembling eq.~(\ref{dipole}). The next order terms, with $m+n=3$,
contribute to higher order moments.  (For spherical symmetry (Rees and
Sciama 1968), only the $m=3$, $n=0$ term is relevant.)  By requiring
contributions at these orders to be smaller than 10$^{-5}$, the level
of anisotropy measured by COBE at $\ga 10^\circ$ angular scale, we can
constrain the size of the deviation $\Delta R$.  For $R\la 100h^{-1}$
Mpc, a bound is estimated by considering the $m=1$, $n=2$ term:
\begin{equation}
\left\vert{\delta T\over T}\right\vert\sim (\Delta H)t_0\left({R\over ct_0}
\right)\left({\Delta R\over R}\right)^2 \la 10^{-5},
\label{m1n2}
\end{equation}
where $\Delta H$ is the difference between the expansion rates inside
and outside the void, which is roughly $H_0(1-\Omega_{\rm void})/3$
in linear theory. Eq.~(\ref{m1n2}) amounts to
\begin{equation}
{\Delta R\over R}\la 0.03\sqrt{{100h^{-1}{\rm Mpc}\over R(1-\Omega_{\rm
void})}}
\end{equation}
at $\ga 10^\circ$ angular scale.
If $R\ga 100h^{-1}$ Mpc, a bound, estimated from the $m=2$, $n=1$
term, gives
\begin{equation}
{\Delta R\over R}\la {0.02\over 1-\Omega_{\rm void}}\left({
100h^{-1}{\rm Mpc}\over R}\right)^2~.
\end{equation}
Therefore, a 200$h^{-1}$ Mpc radius void must be spherically symmetric
to $\sim 1\%$ to avoid a contradiction with the COBE results.

\section{Discussion and Conclusions}

The Hubble constant sets both the distance and time scales for the
Universe and therefore plays a central role in all cosmological
models.  Numerous observations place its value between $65$ and
$95\ksm$.  If indeed the true value is in this range, and if the
Universe is as old as the globular cluster experts say it is, then
fairly radical modifications to the standard $\om=1$ paradigm will be
called for.

In this paper, we explore the possibility that the true Hubble
constant $H_0\simeq 50\ksm$, not because of systematic errors in
current observations, but because these observations do not reach deep
enough into the cosmos.

Possible discrepancies between the locally measured value for $H$ and
the true value are quantified in $\delta_H=\delta_H(R)$ where $R$ is
the radius of the sample volume.  We outline a simple procedure for
calculating $\delta_H$ based on linear perturbation theory.  Where
applicable, we compare the results with those based on the N-body
simulations of TCO.  Our conclusions are as follows:

$\bullet$ Linear and nonlinear theories are in good agreement
on large scales.

$\bullet$ For $R\ga 100\mpc$, $\delta_H$ is insensitive to the type of
dark matter present in the model.

$\bullet$ $\delta_H$ is correlated with the mass fluctuation within the
sample.  In particular, a measurement of $\delta M/M$ leads to the
prediction $\delta H/H_0 = -(0.6 \pm 0.15)\delta M/M$.

$\bullet$ $\delta_H(R=100\mpc)\simeq 0.07$ for standard $n=1$ CDM as well
as $n=1$ MDM.  If these models describe our Universe, then the HST and
CFHT results, in all likelihood, reflect the true value.

$\bullet$ $\delta_H(R=100)\mpc> 0.1$ for nonstandard models with extra
power on large scales.  These include an $n=1$ CDM-like model with
$\Gamma=0.35$ and $\sigma_8=1.2$ (see Eq \EC{cdmps}) and either CDM or
MDM with a tilted, $n=1.5$ primordial spectrum.  If our Universe is
described by either of these models, then measurements of $H$ will
have to reach $\sim 200\mpc$ where $\delta_H$ drops well below $0.1$.

Galaxy counts and redshift surveys allow us the opportunity to
directly probe the density and velocity field in our region of the
Universe.  Unfortunately, the interpretation of these observations is
not so straightforward.  In the case of the galaxy counts, for example,
it is difficult to separate density inhomogeneities from evolutionary
effects.  On the other hand, analysis of the velocity data is made
difficult by sparse sampling.  CBR anisotropy measurements offer an
alternative, and potentially clean probe of the large scale
inhomogeneities discussed in this paper.  Here, we calculate the
expected temperature fluctuations due to voids at the last scattering
surface; between us and the last scattering surface; and by the void
enclosing us.  The results are compared with existing data from the
COBE DMR experiment, and by degree-scale experiments such as MAX and MSAM.
A summary of our conclusions follows:

$\bullet$ No more than a handful of voids with radii $\ga 400\hmpc$
can exist in our Hubble volume.

$\bullet$ Smaller voids are constrained by degree-scale CBR
experiments.  We find that the number density today of voids with radii $\ga
60\hmpc$ and $\Omega_{\rm void}\le 0.55$, under the current sky coverage of
degree-scale CBR experiments, is $\la 3\times 10^{-7} h^3\mpc^{-3}$,
occupying ${\cal O}(0.1)$ of our Hubble volume.

$\bullet$  If the void enclosing us has a radius $\ga 200\hmpc$, large
dipole and quadrupole anisotropies will be induced, unless, of course,
the void is very nearly spherical, and we are near the center.

\section{Acknowledgments}
We thank Andrew Jaffe, Kyle Lake, Man Hoi Lee, Sharon Vadas, and Rick
Watkins for helpful discussions. This work is supported by the Natural
Science and Engineering Research Council of Canada.  X. Shi is
supported, in part, by the Canadian Institute for Theoretical
Astrophysics through the CITA National Fellow program.

\newpage

\section*{References}

\reff Alsop, D. C. et al. 1992, ApJ, 395, 317
\reff Bartlett, J. G. et al. 1994, Nature, submitted; astro-ph/9407061
\reff Bennett, C. L.  et al., 1994, COBE preprint \#94-01; astro-ph/9401012
\reff Berschinger, E. 1985, ApJS, 58, 1
\reff Birkinshaw, M. \& Hughes, J. P., 1994, ApJ, 420, 33
\reff Blaes, O. M., Glodreich, P., \& Villumsen, J. V. 1990, ApJ, 361, 331
\reff Blumenthal, G. R., da Costa, L. N., Goldwirth, D. S., Lecar, M. \&
Piran, T. 1992, ApJ, 388, 234
\reff Broadhurst, T. J. et al. 1990, Nature, 343, 726
\reff Bucher, M., Glodhaber, A. S., \& Turok, N., 1995, hep-ph/9501396
\reff Bunn, E. F., Scott, D. \& White, S. D. M., 1994, astro-ph/9409003
\reff Chaboyer, B. 1994, ApJL, submitted; CITA 94-52
\reff Cheng, E. S. et al. 1994, ApJL, 422, L37
\reff Clapp, A. C. et al. 1994, ApJL, 433, L57
\reff de Lapperent, V., Geller, M. J. \& Huchra, J. P., 1986, ApJL, 302, L1
\reff Devlin, M. J. et al. 1994, ApJL, 430, L1
\reff Dodelson, S. \& Stebbins, A. 1994, ApJ, 433, 440
\reff Dubinski, J., et al., 1993, ApJ, 410, 458
\reff Efsthathio, G., Bond, J.R. \& White S. D. M., 1992, MNRAS, 258, 1p
\reff Fang, L., \& Wu, X. 1993, ApJ, 408, 25
\reff Feldman, H. A., \& Watkins, R. 1994, ApJ, 430, L17
\reff Freedman W. et al. 1994, Nature, 371, 757
\reff Geller, M. J. \& Huechra, J. P., 1989, Science, 246, 897
\reff Gundersen, J. O. et al. 1993, ApJL, 413, L1
\reff Han, M. \& Mould, J. R., 1992, ApJ, 396, 453
\reff Holtzman, J., 1989, ApJSS, 71, 1
\reff Hu, W., Sugiyama, N., \& Silk, J. 1995, Nature, submitted;
astro-ph/9504057.
\reff Jaffe, A. H. \&  Kaiser, N., 1994, ApJL, submitted; astro-ph/9408046
\reff Kaiser, N., 1988, Mon. Not. R. A. S., 231, 149
\reff Kirshner, R. P., Oemeler, A., Schecter, P. L. \& Shectman,
S. A., 1981 ApJL, 248, L57
\reff Kramer, D. et al. 1980, {\sl Exact Solutions of Einstein's
 Field Equations}, Chapter 13, and references therein (Cambridge University
Press).
\reff Lauer, T. R. \& Postman, M., 1994, ApJ, 425, 418
\reff Linde, A. 1995, Phys. Lett. B, 351, 99
\reff Loveday, J., Peterson, B. A., Efstathiou, G., \& Maddox,
S. J. 1992, ApJ 390, 338
\reff Lucchin et al., 1995, ApJ, submitted; astro-ph/9504028
\reff Maddox, S. J., Efstathiou, G., Sutherland, W. J. \&
Loveday, J. 1990, MNRAS, 242, 43p
\reff Maeda, K., \& Sato, H. 1983a, Prog. Theor. Phys., 70, 772
\reff Maeda, K., \& Sato, H. 1983b, {\sl ibid.}, 1276
\reff Mart\'inez-Gonz\'alez, E., Sanz, J. L., \& Silk, J. 1990, ApJL, 355, L5
\reff Meinhold, P., \& Lubin, P. 1991, ApJL, 370, L11
\reff Meinhold, P. et al., 1993, ApJL, 409, L1
\reff Metcalfe, N., Shanks, T., Roche, N., \& Fong, R. 1993, Ann. of
the NY Acad. of Sci., 688, 534
\reff Moffat, J. W., \& Tatarski, D. C. 1994, UTPT-94-19; astro-ph/9407036
\reff Nicolaci da Costa, L. et al., 1991, ApJSS, 75, 935
\reff Peebles, P. J. E. 1993, {\sl Principles of Physical Cosmology}
(Princeton University Press, Princeton, New Jersey)
\reff Pierce, M. et al. 1994, Nature, 371, 385
\reff Piran, T., Lecar, M., Goldwirth, D. S., da Costa, L. N.,
 \& Blumenthal, G. R. 1993, MNRAS, 265, 681
\reff Rees, M. J., \& Sciama, D. W. 1968, Nature, 217, 511
\reff Sachs, R. K., \& Wolfe, A. M. 1968, ApJ, 147, 73
\reff Sandage, A. 1993, AJ, 106, 719
\reff Schmidt, B. P. et al. 1994, ApJ, 432, 42
\reff Shi, X. 1995, ApJ, in press
\reff Smoot, G. et al. 1992, ApJL, 396, L1
\reff Strauss, M. A., et al., 1994 preprint
\reff Suto, Y., Suginohara, T., \& Inagaki, Y. 1994, Prog. Theor. Phys.,
 submitted; astro-ph/9412090
\reff Thompson, K. L., \& Vishniac, E. T. 1987, ApJ, 313, 517
\reff Turner, E. L., Cen, R., \& Ostriker, J. P., 1992, AJ, 103, 1427 (TCO)
\reff van den Bergh, S., 1992, Publ. Astron. Soc. Pac., 104, 861
\reff Wu, X. et al. 1995, ApJL, submitted; astro-ph/9512082
\reff Yamashita, K. 1994 in New Horizon of X-Ray Astronomy,
ed. F. Makino \& T. Ohashi (Universal Academy Press, Tokyo), 279

\newpage

\section*{Figure Captions:}

\noindent Figure 1. $\delta_H$, the rms deviation in the Hubble flow,
as a function of the transfer function parameterized by $\Gamma$ (Eq
\EC{cdmps}) for scales $R=100,\, 150,\,$ and $200\mpc$, $h=0.5$,
$\om=1$, and $\sigma_8=1$.  $\delta_H$ scales linearly with $\sigma_8$.

\bigskip

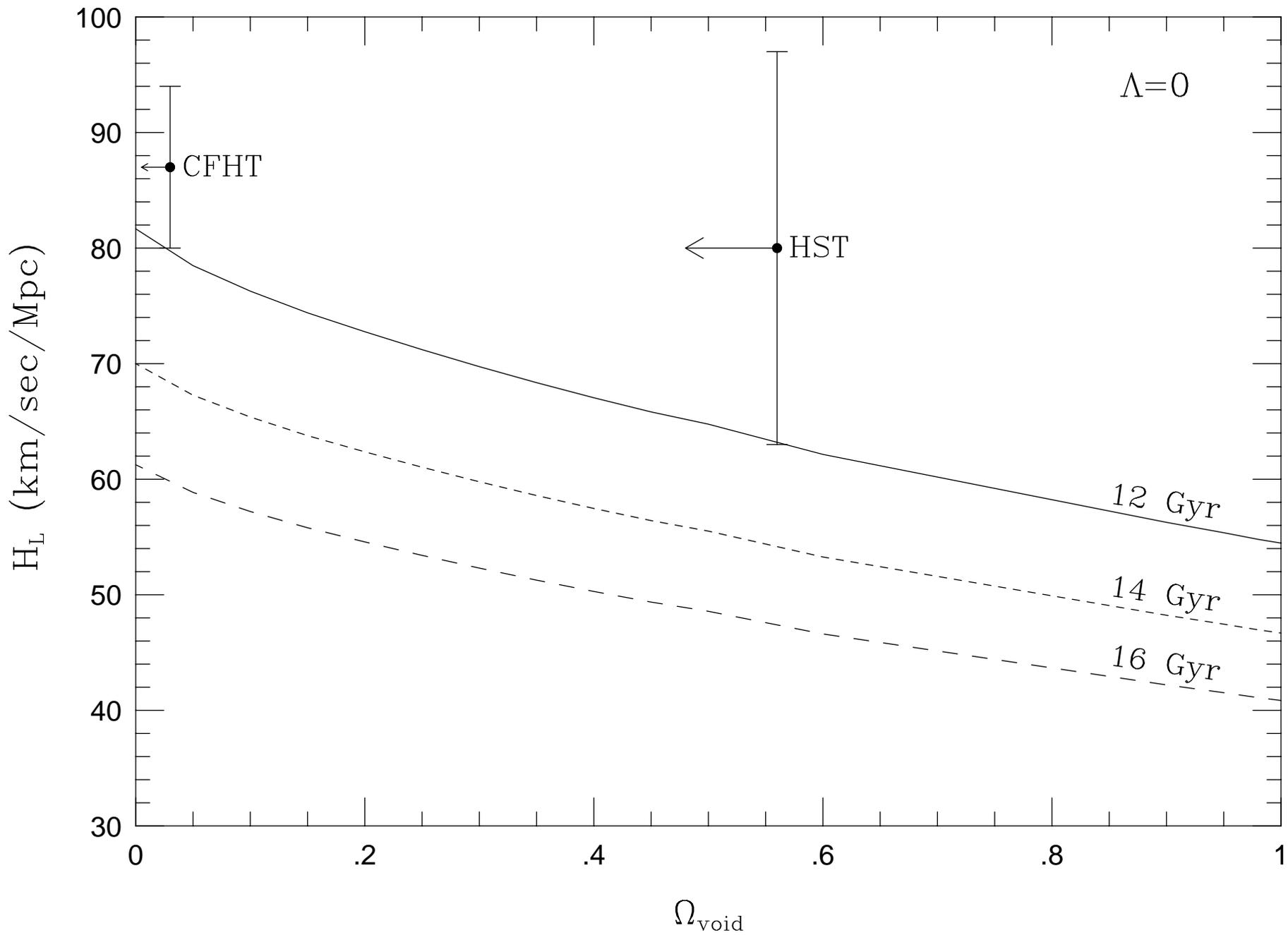
\noindent Figure 2. The relation between $H_L$, $\Omega_{\rm void}$, and $t_0$
in an $\om=1$, $\Lambda=0$ universe.
\bigskip

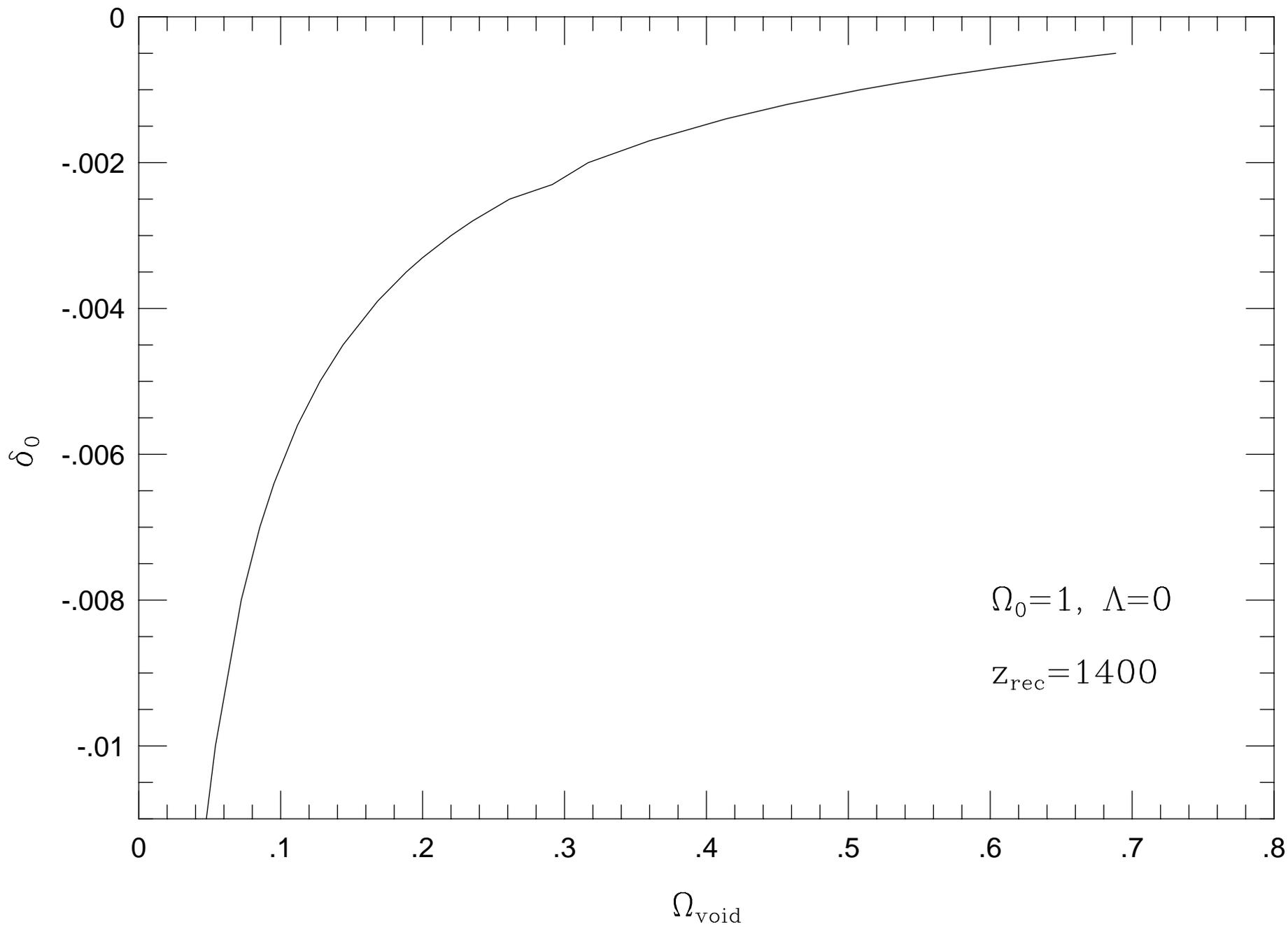
\noindent Figure 3. The relation between the initial density fluctuation
$\dbi$ in a primeval void and the density parameter $\Omega_{\rm void}$
it would have today.
\bigskip

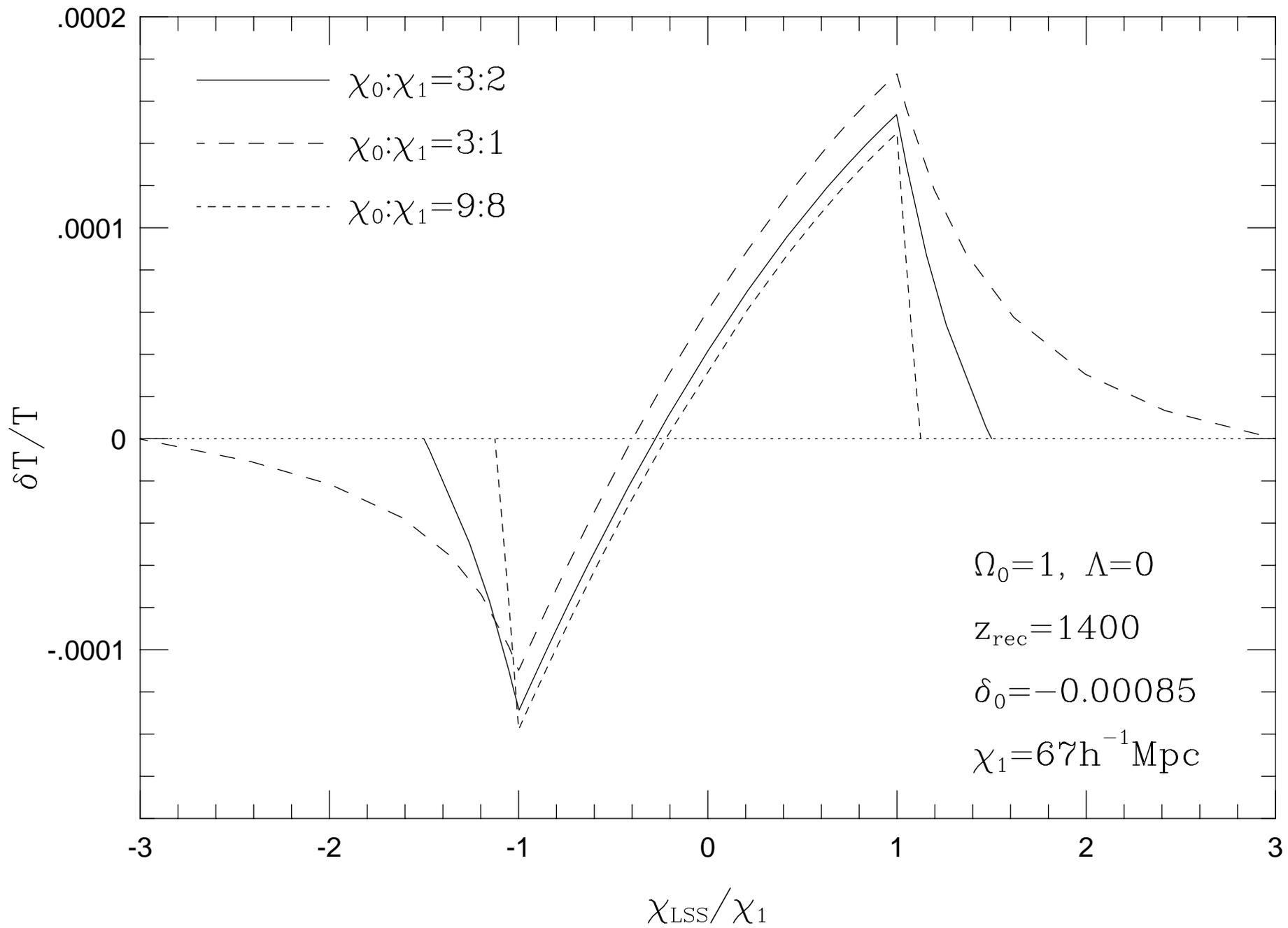
\noindent Figure 4. The temperature fluctuation of CBR in the direction of
the center of a primeval void, due to the Sachs-Wolfe effect, vs.
the position of the last scattering surface relative to the center of
the void. The initial density fluctuation is assumed to be
$\dbi=-8.5\times 10^{-4}$. Different lines stands for cases with
different wall thickness.
\bigskip

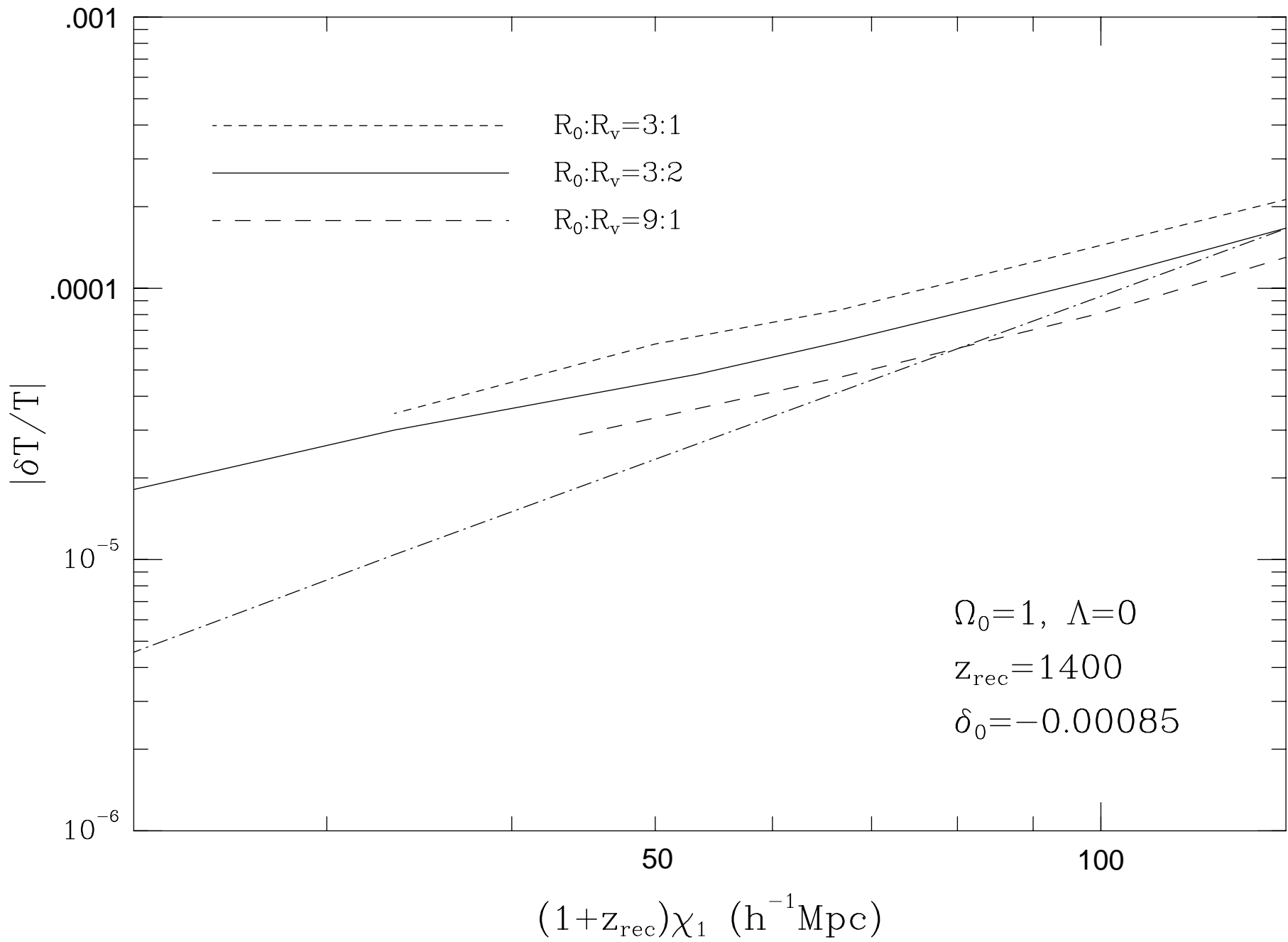
\noindent Figure 5. The averaged temperature fluctuation magnitude
in CBR from a primeval void, vs. the coordinate
radius of the void, for three different wall thickness.
The dot-dash line shows the temperature fluctuation
vs. the coordinate radius of the void when the last scattering surface cut
through the center of the void. The initial density fluctuation is assumed
to be $\dbi=-8.5\times 10^{-4}$.
\bigskip

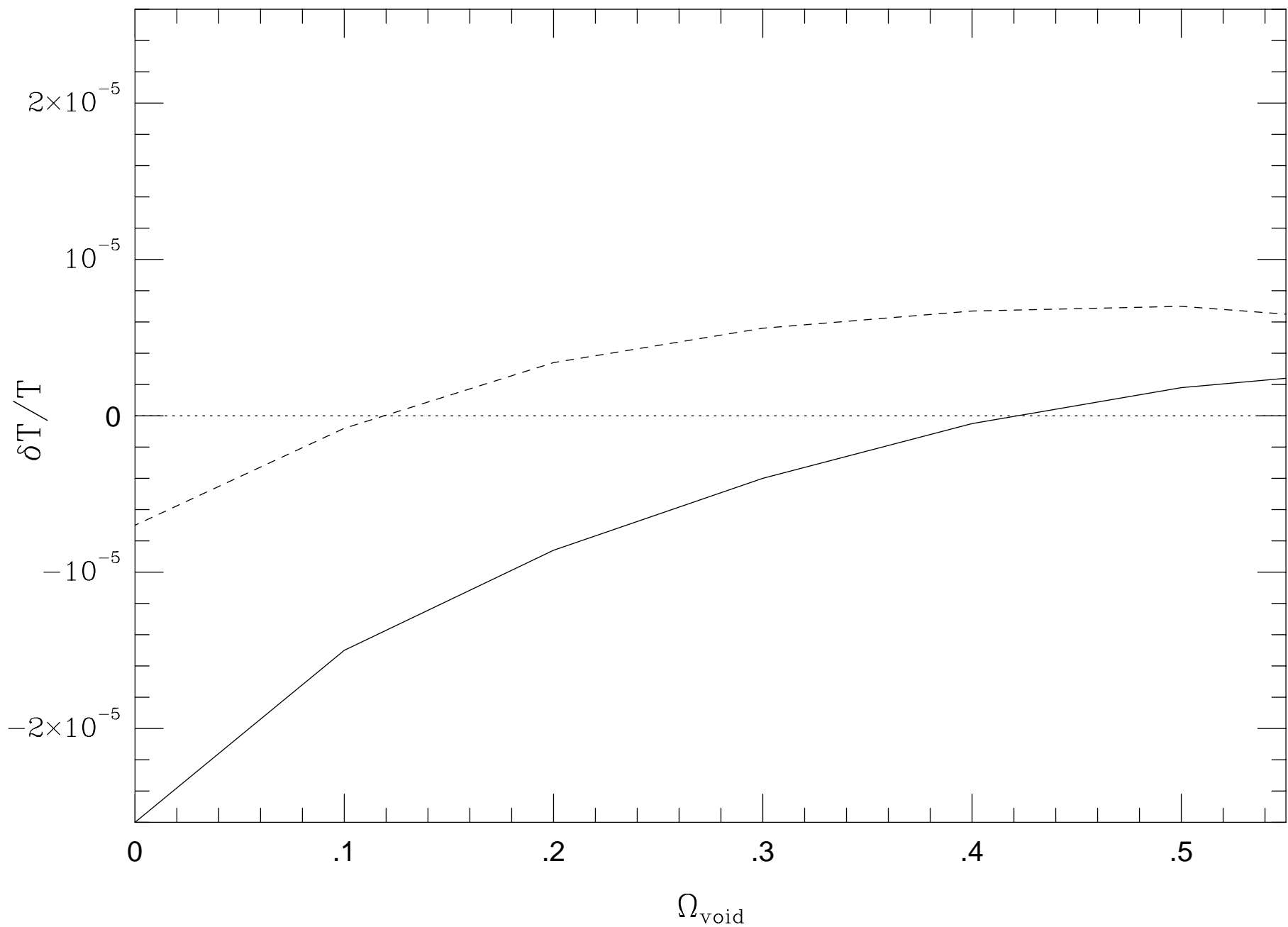
\noindent Figure 6. The temperature fluctuation magnitude
in CBR from a intermediate void with
a radius of 100$h^{-1}$ Mpc, vs. the density parameter $\Omega_{\rm void}$
inside the void. We are assumed to be just outside the void. The solid line:
$\gamma=2/3$; the dash line: $\gamma$ as calculated from momentum conservation
of the wall.
\bigskip

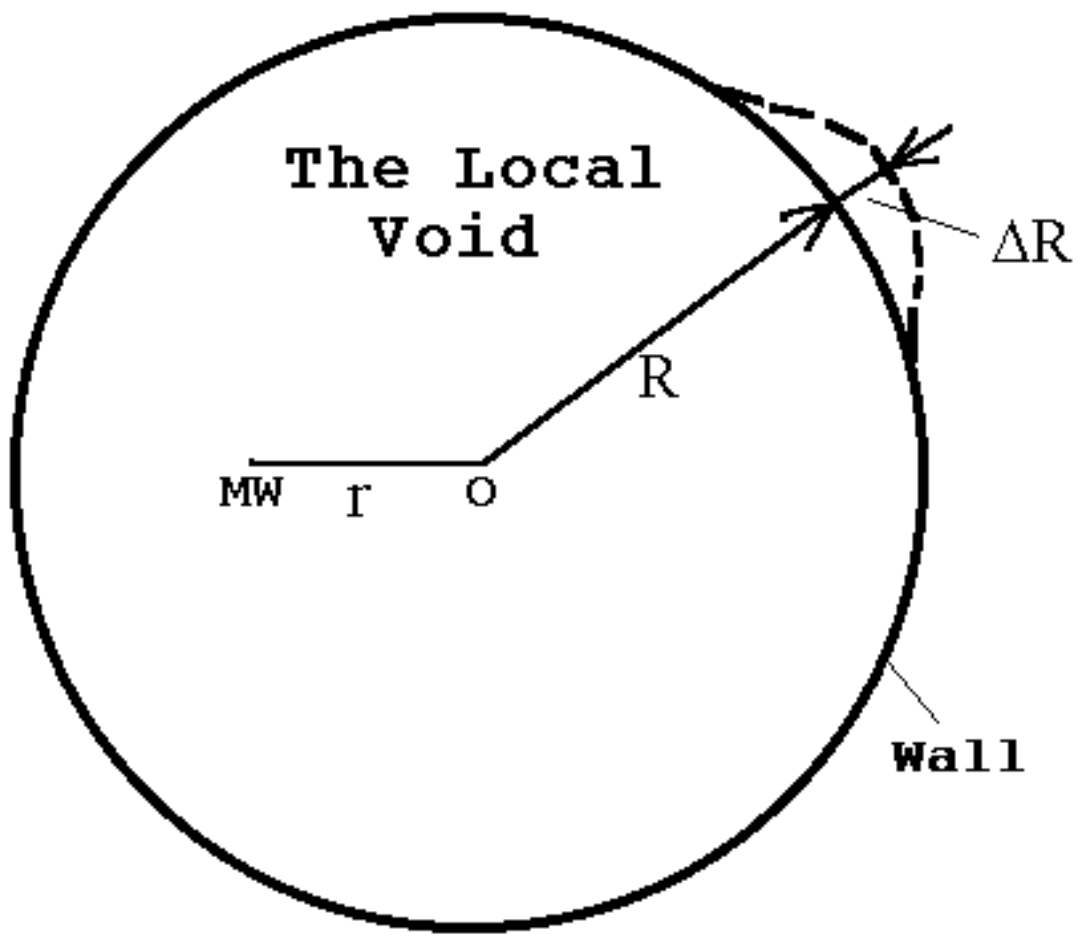
\noindent Figure 7. The geometry of a void enclosing us. The dashed line
illustrate a deviation of the wall of the void from a perfect sphere.
$\Delta R$ is the height of the deviation.
\bigskip

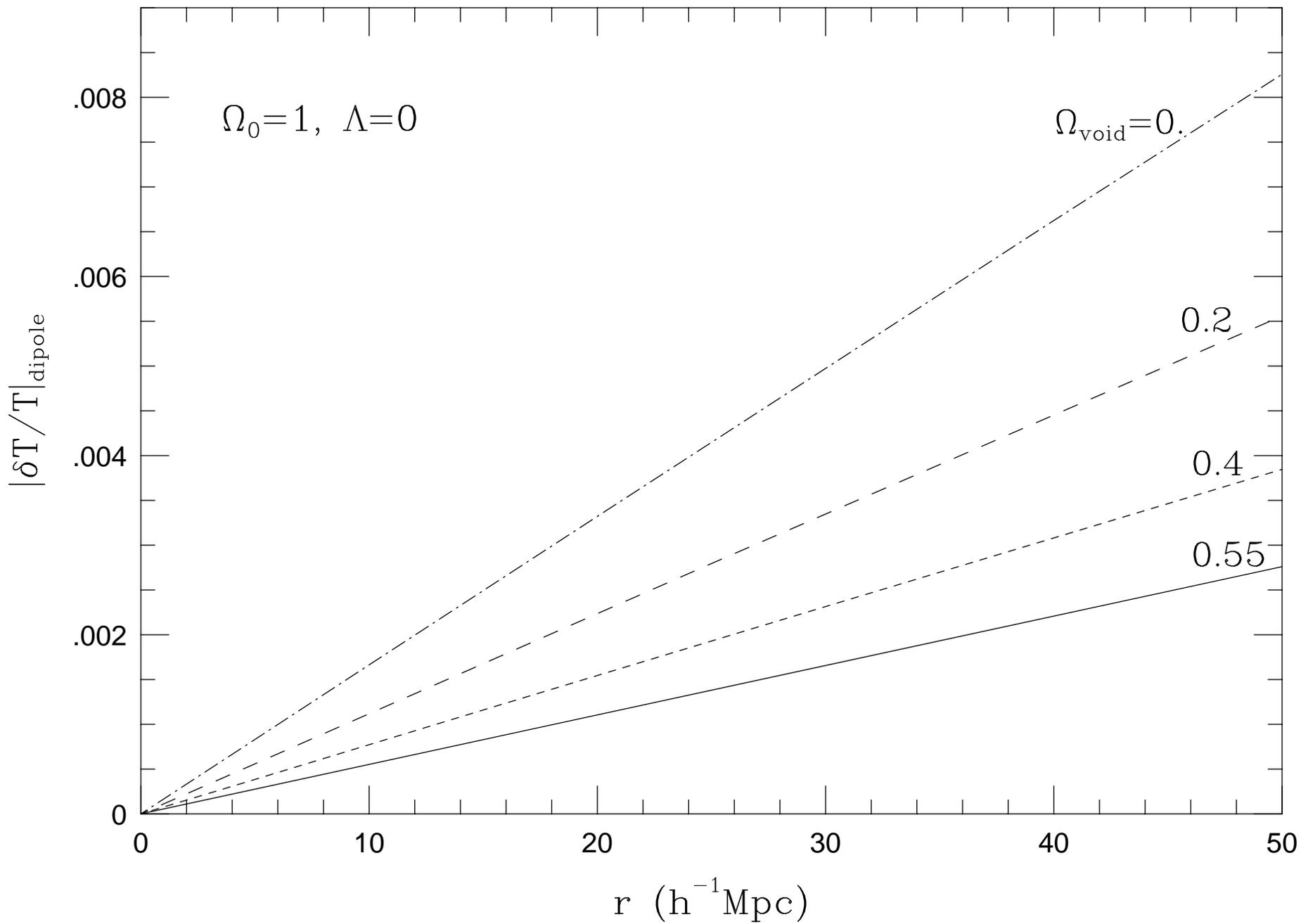
\noindent Figure 8. The amplitude of a CBR dipole moment
$\vert\delta T/T\vert_{\rm dipole}$ contributed by
a void enclosing us vs. the distance between us and the center of the void
$r$. No peculiar velocity other than that due to the faster expansion inside
the void is assumed for us.
\bigskip

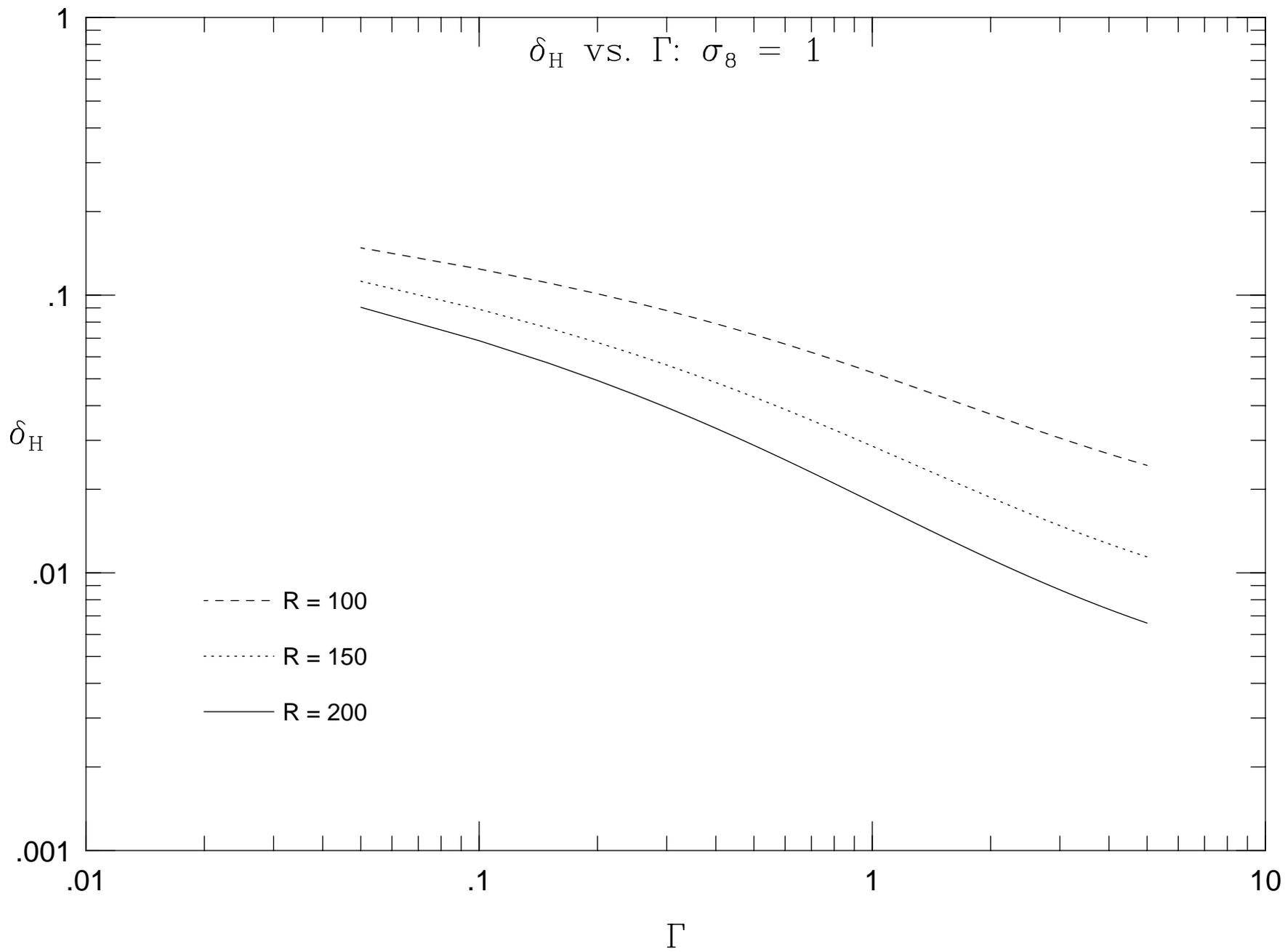
\end{document}